\documentclass[
    prb,
    twoside,
    twocolumn,
    10pt,
    floatfix,
    showpacs,
    citeautoscript,
    superscriptaddress,
]{revtex4-2}

\usepackage{amsmath,amssymb}
\usepackage{booktabs}
\usepackage{dcolumn}
\usepackage{glossaries}
\usepackage{graphicx} 
\usepackage[version=4]{mhchem}
\usepackage{siunitx}
\usepackage[usenames,dvipsnames,svgnames]{xcolor}
\usepackage{hyperref}
\graphicspath{{figures/}}

\hypersetup{
    pdffitwindow=false,
    pdfstartview={FitH},
    pdfnewwindow=true,
    colorlinks=true,
    linkcolor=blue,
    citecolor=blue,
    filecolor=blue,
    urlcolor=blue,
}

\newcolumntype{d}{D{.}{.}{-1}}
\newcolumntype{e}{D{+}{\,\pm\,}{6,5}}
\newcolumntype{f}{D{+}{\,\pm\,}{3,3}}

\renewcommand{\vec}[1]{\ensuremath\boldsymbol{#1}}
\renewcommand{\tensor}[1]{\ensuremath\boldsymbol{#1}}

\setacronymstyle{long-short}

\newacronym{acf}{ACF}{autocorrelation function}
\newacronym{cpu}{CPU}{central processing unit}
\newacronym{dft}{DFT}{density functional theory}
\newacronym{dp}{DP}{deep potential}
\newacronym{gpu}{GPU}{graphics processing unit}
\newacronym{ir}{IR}{infrared}
\newacronym{md}{MD}{molecular dynamics}
\newacronym{ml}{ML}{machine learning}
\newacronym{nep}{NEP}{neuroevolution potential}
\newacronym{pes}{PES}{potential energy surface}
\newacronym{rmse}{RMSE}{root-mean-square error}
\newacronym{rrmse}{RRMSE}{root-mean-square-error relative to standard deviation}
\newacronym{scan}{SCAN}{strongly constrained and appropriately normed}
\newacronym{sagpr}{SA-GPR}{symmetry-adapted Gaussian process regression}
\newacronym{teann}{T-EANN}{tensorial embedded-atom neural network}
\newacronym{tnep}{TNEP}{tensorial neuroevolution potential}

\DeclareSIUnit{\au}{a.u.}
\DeclareSIUnit{\atom}{atom}
\DeclareSIUnit{\electron}{e}
\DeclareSIUnit{\mybohr}{bohr}
\DeclareSIUnit\bar{bar}
\DeclareSIUnit\angstrom{\protect\text{Å}}

\global\let\oldnewlabel\newlabel
\gdef\newlabel#1#2{\newlabelxx{#1}#2}
\gdef\newlabelxx#1#2#3#4#5#6{\oldnewlabel{#1}{{#2}{#3}}}
\let\newlabel\oldnewlabel
\newcommand{\addZhejiangQuzhou}{Institute of Zhejiang University-Quzhou, Quzhou 324000, P.~R. China}
\newcommand{\addZhejiangHangzhou}{College of Chemical and Biological Engineering, Zhejiang University, Hangzhou 310058, P.~R. China}
\newcommand{\addChalmers}{Department of Physics, Chalmers University of Technology, SE-41296, Gothenburg, Sweden}
\newcommand{\addCAS}{State Key Laboratory of Multiphase Complex Systems, Institute of Process Engineering, Chinese Academy of Sciences, Beijing, P.~R. China}
\newcommand{\addWashington}{Department of Chemical Engineering, University of Washington, Seattle, WA 98195, USA}
\newcommand{\addBohai}{College of Physical Science and Technology, Bohai University, Jinzhou 121013, P.~R. China}

\begin{document}

\author{Nan Xu}
\affiliation{\addZhejiangQuzhou}
\affiliation{\addZhejiangHangzhou}

\author{Petter Rosander}
\author{Christian Schäfer}
\author{Eric Lindgren}
\author{Nicklas Österbacka}
\affiliation{\addChalmers}

\author{Mandi Fang}
\affiliation{\addZhejiangQuzhou}
\affiliation{\addZhejiangHangzhou}

\author{Wei Chen}
\affiliation{\addCAS}

\author{Yi He}
\email{yihezj@zju.edu.cn}
\affiliation{\addZhejiangQuzhou}
\affiliation{\addZhejiangHangzhou}
\affiliation{\addWashington}

\author{Zheyong Fan}
\email{brucenju@gmail.com}
\affiliation{\addBohai}

\author{Paul Erhart}
\email{erhart@chalmers.se}
\affiliation{\addChalmers}

\title{
    Tensorial properties via the neuroevolution potential framework:\texorpdfstring{\\}{}
    Fast simulation of infrared and Raman spectra
}

\begin{abstract}
Infrared and Raman spectroscopy are widely used for the characterization of gases, liquids, and solids, as the spectra contain a wealth of information concerning in particular the dynamics of these systems.
Atomic scale simulations can be used to predict such spectra but are often severely limited due to high computational cost or the need for strong approximations that limit application range and reliability.
Here, we introduce a machine learning (ML) accelerated approach that addresses these shortcomings and provides a significant performance boost in terms of data and computational efficiency compared to earlier ML schemes.
To this end, we generalize the neuroevolution potential approach to enable the prediction of rank one and two tensors to obtain the tensorial neuroevolution potential (TNEP) scheme.
We apply the resulting framework to construct models for the dipole moment, polarizability, and susceptibility of molecules, liquids, and solids, and show that our approach compares favorably with several ML models from the literature with respect to accuracy and computational efficiency.
Finally, we demonstrate the application of the TNEP approach to the prediction of infrared and Raman spectra of liquid water, a molecule (\ce{PTAF^-}), and a prototypical perovskite with strong anharmonicity (\ce{BaZrO3}).
The TNEP approach is implemented in the free and open source software package \textsc{gpumd}, which makes this methodology readily available to the scientific community.
\end{abstract}

\maketitle

\section{Introduction}

\Gls{ir} and Raman spectroscopy are widely used techniques for the non-destructive characterization of the dynamics and to some extent chemistry of materials spanning the entire range from the gas phase to condensed matter \cite{bernhard_schrader_infrared_1995, DENDISOVA_2018_ir, DAS_2011_raman}.
Over the years, various theoretical approaches have been developed for simulating \gls{ir} and Raman spectra, including in particular methods based on \textit{ab-initio} \gls{md} simulations \cite{thomas_computing_2013,  Silvestrelli_raman_1997, Putrino_raman_2002, apra_time_2018,Luber2014}.
While these approaches are capable of reproducing experimental \gls{ir} and Raman spectra of gases, liquids and solids \cite{kou_simulating_2020, Luber2014, Silvestrelli_raman_1997, apra_time_2018}, they are severely limited with respect to the system sizes and time scales attainable for two main reasons \cite{Silvestrelli_raman_1997, sommers_raman_2020}:
Firstly, \textit{ab-initio} \gls{md} simulations rely on computationally demanding electronic structure calculations that scale strongly with system size in order to obtain energy and forces at every time step.
Secondly, similarly expensive calculations of dipole moment ($\vec{\mu}$), polarizability ($\tensor{\alpha}$) or electric susceptibility ($\tensor{\chi}$) are required for at least many thousand configurations to achieve numerical convergence of the underlying correlation functions \cite{Silvestrelli_raman_1997}.

\Gls{md} simulations can be accelerated by using classical force fields \cite{CHARMM2000, Jorgensen_opls_1988,  Cornell_amber_1995} or empirical interatomic potentials \cite{EAM_1986, Tersoff_1988}, which approximate the \gls{pes} with physically motivated yet constrained functions and few fitted parameters.
The accuracy of such approaches for general materials is, however, often limited, negatively affecting the prediction of \gls{ir} and Raman spectra \cite{henschel2020theoretical}.
\Gls{ml} potentials are well suited to address this challenge as they bridge between the accuracy of quantum mechanical methods and the computational efficiency of classical force fields or empirical interatomic potentials \cite{Behler_mlff_2016, unke_machine_2021, zhang_end--end_2018, Zhang_2021_PRL, fan2021prb}.
The power of this approach, in particular for capturing vibrational properties of materials has been shown repeatedly, see, e.g., Refs.~\citenum{gastegger_machine_2017, CHMIELA201938, kwac_machine_2021, Shanavas_nnp_2022, fransson_soft_modes_2023}.

The calculation of $\vec{\mu}$, $\tensor{\alpha}$ or $\tensor{\chi}$ can be accelerated using parametric models in similar fashion.
Considering only static charges, the dipole moment is given by $\vec{\mu} = \sum_{i=1}^{N} Q_i \vec{r}_i$, where $Q_i$ and $\vec{r}_i$ are the charge and position of atom $i$.
Many classical force fields \cite{CHARMM2000, Jorgensen_opls_1988, Cornell_amber_1995} assign fixed charges to atoms and thereby provide a convenient approach for calculating $\vec{\mu}$.
Such fixed-charge models neglect, however, polarization effects, which can lead to large errors \cite{leontyev_accounting_2011}.
While this situation can in principle be ameliorated by fluctuating-charge models \cite{thaunay_strategy_2018, giovannini_calculation_2019}, the latter tend to lack robustness and can be difficult to generalize \cite{sommers_raman_2020, Veit_muml_2020}.

Both $\tensor{\alpha}$ and $\tensor{\chi}$ describe the dielectric response to an applied electric field.
For $\tensor{\alpha}$ or $\tensor{\chi}$, the bond polarizability model is one of the most frequently used parametric ones, and has for example been applied to alkanes \cite{smirnov_quantum-chemical_2006, chen_predicting_2017}, zeolites \cite{Bornhauser_zeolite_bpm_2001} as well as carbon nanotubes \cite{Rahmani_nanotube_bpm_2005}.
However, this simple model often suffers from unsatisfactory transferability when used in different environments \cite{bougeard_calculation_2009}.
POLI2VS \cite{hasegawa_polarizable_2011} and MB-pol \cite{Medders2015} are two other parametric models that can be used for predicting $\vec{\mu}$ and $\tensor{\alpha}$, but are limited to molecular systems such as water \cite{sommers_raman_2020}.

The successful applications of \gls{ml} potentials have inspired the development of \gls{ml} dipole, polarizability, and susceptibility models \cite{grisafi_symmetry-adapted_2018, zhang_efficient_2020,  gastegger_machine_2017, Roberto_2020_PRB, MXene_2023_aalto}.
For $\vec{\mu}$, a rank-1 tensor, both partial-charge and the partial-dipole \gls{ml} models have been developed \cite{Veit_muml_2020}.
The objective of the partial-charge models is to assign proper partial charges for atoms in order to fit the total dipole moment \cite{Veit_muml_2020, gastegger_machine_2017, beckmann_infrared_2022}.
Here, one concern is the balance between the fitting quality of $\vec{\mu}$ and the reproducibility of total charges \cite{Veit_muml_2020, gastegger_machine_2017}.
By contrast partial-dipole models such as \gls{sagpr} \cite{grisafi_symmetry-adapted_2018}, \gls{teann}\cite{zhang_efficient_2020}, and \gls{dp} \cite{Roberto_2020_PRB} treat $\vec{\mu}$ as a sum of vectors \cite{Veit_muml_2020, grisafi_symmetry-adapted_2018} that can be determined from atom-centered chemical environments.

While this approach works for $\vec{\mu}$, which is a rank-1 tensor, it does not transfer to the construction of \gls{ml} models for $\tensor{\alpha}$ or $\tensor{\chi}$, which are rank-2 tensors.
This has motivated the pioneering development of the \gls{sagpr} method for tensorial properties \cite{grisafi_symmetry-adapted_2018} as well as later the \gls{teann} \cite{zhang_efficient_2020, Jiang_Bin_jctc_2023} and \gls{dp} models \cite{sommers_raman_2020}. 

The combination of \gls{ml} potentials with \gls{ml} models for $\vec{\mu}$, $\tensor{\alpha}$ or $\tensor{\chi}$ enables the simulations of \gls{ir} and Raman spectra.
This approach has been used to predict, e.g., the \gls{ir} spectra of methanol, n-alkanes, and a peptide \cite{gastegger_machine_2017}, \gls{ir} and Raman spectra of liquid water \cite{sommers_raman_2020, zhang_efficient_2020, Zhang_2021_PRL, zhang_modeling_2021} or the Raman spectra of various solid materials \cite{BerKom23}.
While these earlier studies have established the usefulness of \gls{ml} models for predicting \gls{ir} and Raman spectra, there is still ample room for improvement of current models for $\vec{\mu}$, $\tensor{\alpha}$ or $\tensor{\chi}$ in terms of computational and data efficiency \cite{Veit_muml_2020, zhang_efficient_2020} as well as the accessibility of these techniques in order to lower the threshold for the widespread adoption of such approaches.

\begin{figure}
\centering
\includegraphics[width=0.98\linewidth]{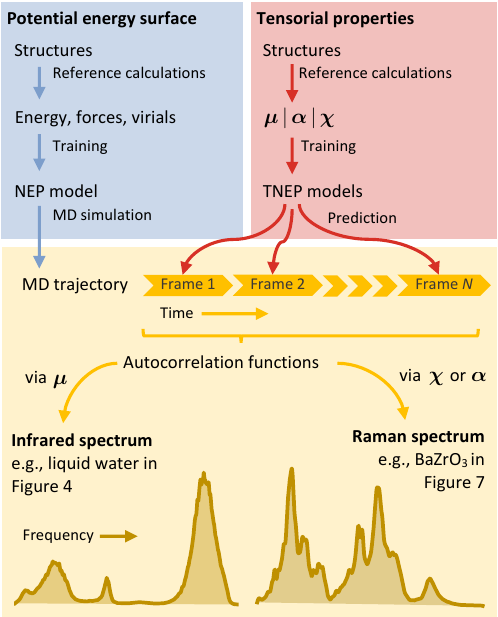}
\caption{
    Workflow for simulations of \gls{ir} and Raman spectra using \gls{nep} models for the \gls{pes} and \gls{tnep} models for the dipole moment $\vec{\mu}$, the polarizability $\tensor{\alpha}$ or the susceptibility $\tensor{\chi}$.
}
\label{fig:workflow}
\end{figure}

This situation motivates the present work, in which we introduce accurate as well as computationally and data efficient \gls{ml} models for rank-1 and rank-2 tensors based on the \gls{nep} framework  \cite{fan2021prb, fan_jpcm_2022, fan_gpumd_2022}.
We demonstrate the efficacy and efficiency of the resulting \gls{tnep} approach by training models for $\vec{\mu}$, $\tensor{\alpha}$, and $\tensor{\chi}$, and combining these with models for the \gls{pes} to predict \gls{ir} and Raman spectra for a molecule (\ce{PTAF^-}), a liquid (water), and a solid (\ce{BaZrO3}; \autoref{fig:workflow}).
We make this methodology available via the \textsc{gpumd} package \cite{fan_gpumd_2022}, enabling comprehensive simulations of high-quality \gls{ir} and Raman spectra with limited user effort.

\section{Methodology}

\subsection{\texorpdfstring{\Gls{nep}}{NEP} models for the \texorpdfstring{\gls{pes}}{PES}}

Since the \gls{ml} models for $\vec{\mu}$ and $\tensor{\alpha}$ that we introduce below are based on the \gls{nep} framework for modeling \glspl{pes} \cite{fan2021prb, fan_jpcm_2022, fan_gpumd_2022}, we first provide a brief review of the latter.
Originally \glspl{nep} are \gls{ml} potentials that model the high-dimensional \gls{pes} of finite or extended systems, in the spirit of the neural network potential model proposed by Behler and Parrinello \cite{Behler2007prl}.
In this formalism, the total energy of the system is given by the sum of atomic site energies $U=\sum_i U_i$.
The site energy $U_i$ for a given atom $i$ depends on the local environment of the atom, which is represented by an abstract vector $q_{i}^{\nu}$ with a number of components indexed by $\nu$.
The function mapping from the descriptor to the site energy is represented by a feedforward neural network (also known as a multilayer perceptron) with typically a single hidden layer.
The input layer of the neural network is thus the descriptor vector and the output layer consists of a single node whose value is the site energy $U_i$ of the considered atom $i$, which can be formally expressed as
\begin{equation}
    U_i = U_i(q_{i}^{\nu}).
    \label{eq:site-energy}
\end{equation}

From the energy, we can derive the rank-2 virial tensor that serves as the foundation for the dipole and polarizability models developed in the present work.
For a given structure with $N$ atoms, the virial tensor can be expressed as \cite{fan_gpumd_2022}
\begin{align}
    W^{\upsilon\nu} = - \sum_i^N \sum_{j \neq i} r_{ij}^{\upsilon} \frac{\partial U_i}{\partial r_{ij}^{\nu}},
    \label{eq:virial}
\end{align}
where $r_{ij}^{\upsilon}$ is the $\upsilon$-component of the vector $\vec{r}_{ij} \equiv \vec{r}_j - \vec{r}_i$, and $\vec{r}_{i}$ is the position of atom $i$.
We refer to the term $\partial U_i / \partial r_{ij}^{\nu}$ as the partial force, explicit expressions for which have been presented in the original works developing the \gls{nep} approach \cite{fan2021prb, fan_gpumd_2022}. 

\subsection{\texorpdfstring{\Gls{tnep}}{TNEP} rank-1 tensor models}

To develop a \gls{ml} model for predicting $\vec{\mu}$, we first note that it is a rank-1 tensor commonly expressed as a vector, in contrast to the energy, which is a rank-0 tensor (i.e., a scalar).
The partial force in Eq.~\eqref{eq:virial} is a vector, but the summation of it over the whole structure would be zero as a result of Newton's third law.
To obtain a vector representation that does not vanish for a general structure, we note that the quantity defined in Eq.~\eqref{eq:virial} is a rank-2 tensor that can adopt both positive and negative values (as it is the virial tensor in the context of \gls{pes} models).
We can thus obtain an expression for a vector quantity by contracting this rank-2 tensor with a vector.
A natural choice for the vector to be contracted is $\vec{r}_{ij}$, which yields the following expression for rank-1 tensors such as the dipole moment
\begin{align}
    \label{equation:mu}
    \vec{\mu} &= - \sum_i^N \sum_{j\neq i} \vec{r}_{ij} \cdot 
    \left(\vec{r}_{ij} \otimes \frac{\partial U_i}{\partial \vec{r}_{ij}}\right) \nonumber \\
    &= - \sum_i^N \sum_{j\neq i} r_{ij}^2 
    \left(\frac{\partial U_i}{\partial \vec{r}_{ij}}\right),
\end{align}
where $r_{ij}^2 = \vec{r}_{ij} \cdot \vec{r}_{ij}$ is the distance squared between atoms $i$ and $j$.
We note that $U_i$ here should have the dimension of charge instead of energy.
Crucially this goes to show that the \gls{nep} formalism for \glspl{pes} can be directly used to construct a \gls{ml} model for rank-1 tensors such as the dipole moment.
Below we refer to Eq.~\eqref{equation:mu} as the \gls{tnep} dipole model. 

\subsection{\texorpdfstring{\Gls{tnep}}{TNEP} rank-2 tensor models}

To develop \gls{ml} models for predicting $\tensor{\alpha}$ or $\tensor{\chi}$, we first note that these are rank-2 tensors.
Clearly, the quantity defined in Eq.~\eqref{eq:virial} is an ideal candidate.
However, using only Eq.~\eqref{eq:virial} to represent $\tensor{\alpha}$ or $\tensor{\chi}$ does not lead to high regression accuracy because the diagonal terms of $\tensor{\alpha}$ or $\tensor{\chi}$ are usually much larger than the off-diagonal ones.
We therefore represent $\tensor{\alpha}$ (and equivalently $\tensor{\chi}$) as a combination of Eqs.~\eqref{eq:site-energy} and \eqref{eq:virial} as follows
\begin{align}
    \alpha^{\upsilon\nu} = \sum_i^N U_i\delta^{\upsilon\nu} - \sum_i^N \sum_{j \neq i} r_{ij}^{\upsilon} \frac{\partial U_i}{\partial r_{ij}^{\nu}},
    \label{eq:polarizability-model}
\end{align}
where $\delta^{\upsilon\nu}$ is the Kronecker delta.
Note that both the first and second term on the right-hand side contribute to the diagonal elements of $\alpha^{\upsilon\nu}$, but only the second term contributes to the off-diagonal elements.
$U_i$ here has the dimension of polarizability instead of energy, yet the entire \gls{nep} formalism can be reused.
Below we refer to Eq.~\eqref{eq:polarizability-model} as the \gls{tnep} polarizability or susceptibility model.

\subsection{Loss functions}

The \gls{nep} approach is named after the underlying \gls{ml} model (a neural network) and the separable natural evolution strategy used as the training algorithm \cite{Schaul_nes_2011}.
The latter is a principled real-valued black-box optimization method that is very well suited for training the weight and bias parameters in the neural network, of which there are typically a few thousand.
The optimization is driven by the minimization of a loss function that is given by the weighted sum of the \glspl{rmse} of physical quantities as well as $\mathcal{L}_1$ and $\mathcal{L}_2$ regularization terms.
For the construction of \gls{pes} models, the physical quantities included in the loss function are the energies, forces, and virial tensors of the structures in the training set,
\begin{align}
L(\vec{z}) =& \lambda_\text{e} \Delta U(\vec{z}) + \lambda_\text{f} \Delta F(\vec{z}) + \lambda_\text{v} \Delta W(\vec{z}) \nonumber \\
    &+ \text{regularization~terms},
\end{align}
where $\Delta U(\vec{z})$, $\Delta F(\vec{z})$, and $\Delta W(\vec{z})$ are the \glspl{rmse} of energies, forces, and virials calculated using a set of trainable parameters $\vec{z}$, and $\lambda_\text{e}$, $\lambda_\text{f}$, and $\lambda_\text{v}$ are the corresponding relative weights.
Explicit expressions for the regularization~terms can be found in Ref.~\citenum{fan_gpumd_2022}.
For the construction of  dipole \gls{tnep} models, the loss function is defined in terms of the \gls{rmse} of the dipole $\Delta \mu(\vec{z})$
\begin{equation}
L(\vec{z}) = \Delta \mu(\vec{z})
    + \text{regularization~terms}.
\end{equation}
For the construction of polarizability \gls{tnep} models, the loss function is defined in terms of the \gls{rmse} of the polarizability $\Delta \alpha(\vec{z})$
\begin{equation}
L(\vec{z}) = \Delta \alpha(\vec{z})
    + \text{regularization~terms}.
\end{equation}

\subsection{Dielectric response}
\label{sect:dielectric-response}

It is instructive to recall some relations that describe the response of finite (such as molecules) and extended systems (such as solids and liquids) to an applied electric field.

If a molecule is subjected to an electric field $\vec{E}$ the resulting displacement of nuclei and electrons induces a dipole, which is given by \cite{mcquarrie_book_1976}
\begin{equation*}
    \boldsymbol{\mu}_\text{ind} = \tensor{\alpha}\boldsymbol{E},
\end{equation*}
where $\tensor{\alpha}$ is the \emph{molecular polarizability}.

For an extended system such as a solid or a liquid, one considers equivalently the dipole moment per unit volume, i.e., the polarization
\begin{equation*}
    \boldsymbol{P} = \epsilon_0\tensor{\chi}\boldsymbol{E},
\end{equation*}
where $\tensor{\chi}$ is the \emph{electric susceptibility}.
In the context of bulk liquids the latter has also been referred to as the bulk polarizability.
For clarity in the following, we use the term polarizability only to refer to the molecular polarizability.
There are different conventions for expressing $\tensor{\mu}$, $\tensor{\alpha}$, and $\tensor{\chi}$ leading to different units (\autoref{sect:note-on-units}).
Here, we use \unit{\electron\cdot\mybohr} for $\vec{\mu}$ and \unit{\mybohr \cubed} for $\tensor{\alpha}$ whereas $\tensor{\chi}$ is unitless.

We note that under certain conditions, one can approximately connect the molecular polarizability and the electric susceptibility via the Clausius-Mossotti relation, which is based on a mean-field treatment of local field effects (see \autoref{sect:clausius-mossotti-relation} in the Supporting Information). 

\subsection{The IR intensity}

The \gls{ir} absorption cross section is given by \cite{mcquarrie_book_1976}
\begin{align}
    \sigma(\omega) = \frac{4\pi^2}{\hbar c n}\omega\left(1 - e^{-\beta\hbar\omega}\right) M(\omega),
    \label{eq:ir-intensity-full}
\end{align}
where $n$ is the refractive index of the material, $c$ the speed of light, $\beta = 1/k_\text{B} T$ and $M(\omega)$ is the absorption lineshape given by the Fourier transform of the \gls{acf} of the (total) dipole moment $\vec{\mu}$,
\begin{align*}
    M\left( \omega \right) = \frac{1}{2\pi} \int_{-\infty}^{\infty}
    \left< \left(\hat{\vec{\epsilon}}\cdot\vec{\mu}\left( 0 \right)\right) \left(\hat{\vec{\epsilon}}\cdot \vec{\mu}\left( t \right)\right) \right>
    e^{-i\omega t}\text{d}t,
\end{align*}
where $\left< \cdots \right>$ indicates the average over time origins and $\hat{\vec{\epsilon}}$ is the polarization of the light \cite{mcquarrie_book_1976}.
For an isotropic sample, the time correlation should be averaged over the three directions, i.e., the lineshape reduces to one third of the trace of the dipole time correlation.
Since the lineshape is sampled classically, we make a classical approximation for the prefactor by expanding the Boltzmann factor to first order, which gives
\begin{align}
    \sigma(\omega) \propto \omega^2 M(\omega).
    \label{eq:ir-intensity}
\end{align}

\subsection{The Raman intensity}

The differential Raman cross-section for Stokes scattering is given by \cite{mcquarrie_book_1976, Cardona_1983, Born_Huang_1954}
\begin{align}
    \frac{\partial^2\sigma}{\partial\omega_\text{out}\partial\Omega} = \left(\frac{\omega_\text{in} - \omega}{c}\right)^4\sum_{\gamma\delta\mu\nu}\hat{n}_{\gamma}\hat{n}_{\mu}L_{\gamma\delta\mu\nu}(\omega)\hat{\epsilon}_{\delta}\hat{\epsilon}_{\nu},
    \label{eq:raman-cross-section}
\end{align}
where $\hat{\vec{n}}$ is the polarization of observed light, $\hat{\vec{\epsilon}}$ is the polarization of the incoming light, and $\Omega$ is a solid angle.
Here, it is assumed that the frequency of the incoming light $\omega_\text{in}$ is significantly larger than the Raman shift $\omega$, and significantly smaller than the band gap, i.e., far from any electronic excitations.
$\tensor{L}(\omega)$ is the Raman lineshape given by the Fourier transform of the time-dependent polarizability $\tensor{\alpha}(t)$ (finite systems) or susceptibility $\tensor{\chi}(t)$ (extended systems), e.g., in the case of the former
\begin{equation}   
    \label{eq:raman-lineshape}
    \begin{split}
         L_{\gamma\delta\mu\nu}(\omega) = \frac{1}{2\pi}\int_{-\infty}^{\infty}&\left< \alpha_{\gamma\delta}\left( 0 \right)\alpha_{\mu\nu}\left( t \right)\right>e^{-i\omega t}\text{d}t.
    \end{split}
\end{equation}
Note that the elements of the polarizability (or susceptibility) tensor are selected by the polarization of the incoming and outgoing light as indicated in Eq.~\eqref{eq:raman-cross-section}.
Polarized Raman measurements can be directly related to Eq.~\eqref{eq:raman-cross-section} by combinations of the Raman lineshape $\tensor{L}(\omega)$.
One can also calculate an average spectrum for isotropic samples \cite{mcquarrie_book_1976}.
The polarizability tensor (and equivalently the susceptibility tensor) can also be written as $\tensor{\alpha} = \gamma\tensor{I} + \tensor{\beta}$ where $\gamma=\text{Tr}(\tensor{\alpha})/3$ and $\tensor{\beta}$ is a traceless tensor to obtain the isotropic (polarized) and anisotropic (depolarized) spectrum.
This leads to the decomposition
\begin{align}
\begin{split}
    L_\text{iso}(\omega)
    &\propto \int_{-\infty}^{\infty}{\left< \gamma\left( 0 \right) \gamma\left( t \right) \right>}e^{-i\omega t}\text{d}t
    \\
    L_\text{aniso}(\omega)
    &\propto \int_{-\infty}^{\infty}{\left< \text{Tr[}\vec{\beta }\left( 0 \right) \vec{\beta }\left( t \right) ] \right>}e^{-i\omega t}\text{d}t.
\end{split}
\label{eq:raman-lineshape-isotropic-anisotropic}
\end{align}

The electric susceptibility (\autoref{sect:dielectric-response}) can be separated into an electronic and an ionic contribution
\begin{equation*}
    \tensor{\chi} = \tensor{\chi}_\text{ion}(\omega) + \tensor{\chi}_\text{e}(\omega),
\end{equation*}
where the general frequency dependence of these terms is emphasized.
For the prediction of Raman spectra we only need to consider the electronic contribution $\tensor{\chi_\text{e}}(\omega)$.
Furthermore, we limit ourselves to non-resonant Raman spectroscopy.
This means that we require the electric susceptibility in the ion-clamped static limit, i.e., $\tensor{\chi}_\text{e}(0)$, and do not have to consider the frequency dependence of $\tensor{\chi}_\text{e}(\omega)$, which arises from electronic transitions.

\subsection{Workflow for simulations of IR and Raman spectra}
\label{sect:workflow}

By combining a \gls{nep} model for the \gls{pes} with \gls{tnep} models for dipole, polarizability or susceptibility, one obtains a simple yet general workflow for the computation of \gls{ir} and Raman spectra (\autoref{fig:workflow}).
Starting from a \gls{nep} \gls{pes} model, large-scale \gls{md} simulations are performed to sample the \gls{pes} via the \textsc{gpumd} package, typically for a few hundred picoseconds.
\Gls{tnep} dipole, polarizability or susceptibility models are then employed to predict $\vec{\mu}(t)$, $\tensor{\alpha}(t)$ or $\tensor{\chi}(t)$ along the trajectory.
Finally, \gls{ir} or Raman spectra are obtained via Fourier transformation of the respective \glspl{acf} via Eqs.~\eqref{eq:ir-intensity} or \eqref{eq:raman-cross-section}.

\section{Performance evaluation}
In this section, we evaluate the performance of \gls{tnep} dipole, polarizability, and susceptibility models in comparison with models from the literature with respect to both regression accuracy and computational speed.
The comparison includes the molecules \ce{H2O}, \ce{(H2O)2}, and \ce{H5O2^+} (the Zundel cation), as well as a set of configurations representing liquid water.
Structures with dipole, polarizability, and/or susceptibility data were retrieved from the repository maintained by the developers of the \gls{sagpr} models \cite {tensoap, grisafi_symmetry-adapted_2018} (see \autoref{sect:dipole-polarizability-data-water} in the Supporting Information for details).
The data set for each of these systems comprises \num{1000} configurations, half of which were use for training, while the other half were used for validation.
The hyperparameters used in the training of the \gls{tnep} models are presented in Tables \ref{tab:model-hyperparameters-dipole-models} and \ref{tab:model-hyperparameters-polarizability-models}.
In the case of the \gls{sagpr} method, the results for liquid water were computed using a publicly available model \cite{tensoap_fast} while the models for the molecules were trained by us (see \autoref{sect:training-of-sagpr-models} for details).
In the case of the \gls{teann} method, we only use those data available in the literature \cite{zhang_efficient_2020}.

\subsection{Dipole moment}
\label{sect:dipole-model}

\begin{figure}
\centering
\includegraphics{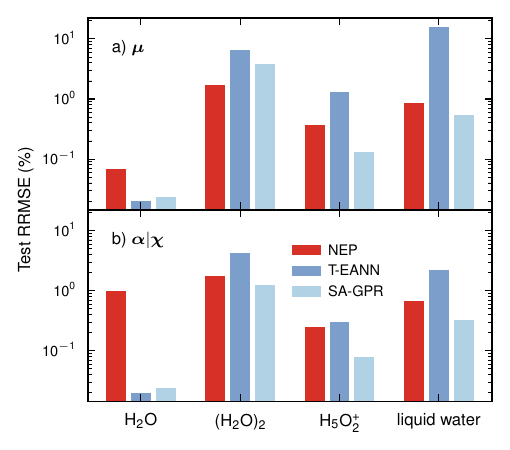}
\caption{
    \acrshortpl{rrmse} for the validation sets according to \gls{tnep}, \gls{teann}, and \gls{sagpr} models for water systems for (a) $\vec{\mu}$ as well as (b) $\tensor{\alpha}$ and $\tensor{\chi}/\rho$.
    Validation \acrshortpl{rrmse} for liquid water from \gls{teann} \cite{zhang_efficient_2020} was reported for the averaged molecular polarizability obtained via the Clausius-Mossotti relation (\autoref{sect:clausius-mossotti-relation}).
    The validation \acrshortpl{rrmse} for $\tensor{\chi}/\rho$ should be somewhat higher than that for the averaged molecular polarizability (also see \autoref{tab:raw-validation-data-water}).
}
\label{fig:test-errors-water}
\end{figure}

The \gls{tnep} dipole models can achieve very high precision when predicting $\vec{\mu}$ for both molecules and liquid water with very low \glspl{rmse} (\autoref{tab:tb1}) and coefficients of determination ($R^2$) very close to one (\autoref{fig:water_dipole_si}).

\begin{table}
    \centering
    \caption{
        \Glspl{rmse} (in \unit{\electron\cdot\mybohr}) and \acrshortpl{rrmse} (unitless) for $\vec{\mu}$ for the validation sets using \gls{nep} rank-1 tensor models.
        For liquid water, the dipole moment is given per water molecule.
    }
    \begin{tabular}{lrc}
    \toprule
    System & \multicolumn{1}{c}{\acrshort{rmse}} & \multicolumn{1}{c}{\acrshort{rrmse}}
    \\
    \midrule
    \ce{H2O}     & \num{2e-4}   & 0.069\%\\
    \ce{(H2O)2}  & \num{105e-4} & 1.681\%\\
    \ce{H5O2^+}  & \num{14e-4}  & 0.371\%\\
    liquid water & \num{17e-4}  & 0.852\%\\
    \bottomrule
    \end{tabular}
    \label{tab:tb1}
\end{table}

As a further, more intuitive measure, one can also consider the \gls{rrmse} \cite{zhang_efficient_2020}, defined as the \gls{rmse} divided by the standard deviation of the reference data (\autoref{fig:test-errors-water}a).
For the water monomer (\ce{H2O}) all three methods yield extremely small \glspl{rrmse} below 0.1\%.
For the other three systems, including liquid water, the \gls{tnep} and \gls{sagpr} models achieve comparable accuracy while the \gls{teann} models perform systematically worse.
This behavior is particularly pronounced for liquid water and might arise since the \gls{teann} model uses the positions relative to the center of mass as input, which are not well defined in periodic systems \cite{Makov_pbc_1995, Luber_dipole_2023}.

\textbf{Neutral molecules}.
The $\vec{\mu}$ of neutral molecules such as \ce{H2O} or \ce{(H2O)2} is uniquely defined.
In the \gls{tnep} approach $\vec{\mu}$ is calculated by summing over atomic contributions which, by contrast to, e.g., the \gls{teann} approach, does not require choosing a reference point.
Therefore, the \gls{tnep} dipole models are naturally suitable for neutral molecules.

In this context, we note that we also trained and validated a model for the QM7B data set containing thousands of neutral organic molecules \cite {Rupp_qm7b_2012, yang_quantum_2019}, for which we make similar observations (\autoref{sect:models-for-qm7b}).
The \gls{tnep} model yields a very low \gls{rmse} for the validation set of \qty{1.80e-3}{\unit{\electron\cdot\mybohr\per\atom}} and a very high $R^2$ score for the validation set of about \num{0.998}.

\textbf{Charged molecules}.
The $\vec{\mu}$ of charged molecules is non-unique and depends on the choice of the reference point \cite{thomas_computing_2013, dipole_book}.
For charged molecules, one should therefore employ the relative permanent dipole $\vec{\mu}_r$ defined with respect to the center of mass, when training \gls{tnep} dipole models.
The reference $\vec{\mu}$ in the \ce{H5O2^+} data set \cite {tensoap, grisafi_symmetry-adapted_2018} have already been transformed to $\vec\mu_r$.
Therefore, the absolute dipole moment of \ce{H5O2^+} including the movement of the center of mass should then be $\vec{\mu} = \vec{\mu}_r + e\cdot \vec{r}_{\text{COM}}$.
The same procedure has been applied to the \ce{PTAF^-} molecule below (\autoref{sect:ptaf}).

\textbf{Periodic systems}.
Traditional methods for calculating $\vec{\mu}$ cannot be applied to periodic systems since the position operator is not uniquely defined \cite{Luber_dipole_2023, SPALDIN_polarization_2012}.
This issue is overcome via the modern theory of polarization \cite{SPALDIN_polarization_2012, grisafi_symmetry-adapted_2018, Krishnamoorthy_water_epsilon_2021}, which provides a rigorous definition for the polarization of periodic systems and established a methodology for calculating $\vec{\mu}$.
It was therefore used in the present work to obtain $\vec{\mu}$ for periodic systems including water (\autoref{sect:dipole-polarizability-data-water}) and $\alpha$-\ce{Fe2O3} (\autoref{sect:dipole-data-for-iron-oxide}).The \gls{tnep} model for $\alpha$-\ce{Fe2O3} yields a very high $R^2$ score for the validation set close to one.

\subsection{Polarizability and susceptibility}
\label{sect:polarizability-model}

The \glspl{rmse} for the diagonal and off-diagonal elements of $\tensor{\alpha}$ of \ce{(H2O)}, \ce{(H2O)2} and \ce{H5O2^+} are quite small (\autoref{tab:tb2}), indicating the high accuracy of the \gls{tnep} polarizability model.
The coefficients of determination are larger than 0.98 mirroring this trend (\autoref{fig:water_polar_diag_si} and \autoref{fig:water_polar_offdiag_si}).
For liquid water, we consider $\tensor{\chi}/\rho$, which has the same unit as the polarizability per atom.
The \glspl{rmse} for $\tensor{\chi}/\rho$ are on the same order of magnitude as the \glspl{rmse} for $\tensor{\alpha}$ (\autoref{tab:tb2}).

The \gls{nep} models achieve an accuracy that is comparable to the \gls{teann} and \gls{sagpr} models for the polarizability of \ce{(H2O)2} and \ce{H5O2^+} as well as the susceptibility of liquid water (\autoref{fig:test-errors-water}b).
While the performance for the water monomer \ce{H2O} is worse, the \gls{tnep} model still yields a validation \gls{rrmse} of less than 1\%.

As a further test we constructed a \gls{tnep} polarizability model for the QM7B data set (\autoref{sect:models-for-qm7b}).
The \gls{rmse} values for the validation set are \qty{4.64e-2}{\mybohr \cubed\per\atom} and \qty{2.58e-2}{\mybohr\cubed\per\atom} for the diagonal and off-diagonal elements of $\tensor{\alpha}$, respectively.
For comparison, Wilkins \textit{et al.} \cite{wilkins_accurate_2019} reported a higher \gls{rmse} value of \qty{5.50e-2}{\mybohr \cubed\per\atom} over both the diagonal and off-diagonal elements of $\tensor{\alpha}$ using a \gls{sagpr} model.

\begin{table}
    \centering
    \caption{
        \Glspl{rmse} (in \unit{\mybohr\cubed}) and \glspl{rrmse} (unitless) for $\tensor{\alpha}$ (molecules) and $\tensor{\chi}/\rho$ (liquid water) for the validation sets using \gls{tnep} rank-2 tensor models.
        For liquid water, the $\tensor{\chi}/\rho$ is given per water molecule.
    }
    \label{tab:tb2}

    \begin{tabular}{crdrd}
     \toprule
      & \multicolumn{2}{c}{diagonal elements} & \multicolumn{2}{c}{off-diagonal elements}\\
     \cmidrule(lr){2-3}   \cmidrule(lr){4-5}
    System & \multicolumn{1}{c}{\acrshort{rmse}} & \multicolumn{1}{c}{\acrshort{rrmse}} &  \multicolumn{1}{c}{\acrshort{rmse}} & \multicolumn{1}{c}{\acrshort{rrmse}} \\
    \midrule
    \ce{H2O}     & \num{85e-3}  &  5.89\%  & \num{4e-3}   &  1.22\% \\
    \ce{(H2O)2}  & \num{227e-3} &  8.82\%  & \num{137e-3} & 12.59\% \\
    \ce{H5O2^+}  & \num{23e-3}  &  1.20\%  & \num{17e-3}  &  1.06\% \\
    liquid water & \num{54e-3}  & 16.28\%  & \num{37e-3}  & 20.38\% \\
    \bottomrule
    \end{tabular}%  
\end{table}

\subsection{Computational speed}
It is now instructive to evaluate the computational performance of \gls{tnep} models in comparison with publicly available \gls{sagpr} models \cite{tensoap, tensoap_fast}.
To this end, we consider liquid water systems with varying numbers of atoms.
Starting from a cell containing 96 atoms, larger samples with up to \num{69984} atoms were created by replication.

The \gls{sagpr} models can only be run serially on a \gls{cpu}.
In contrast, the \gls{tnep} model can be run on \glspl{cpu} using \texttt{NEP\_CPU} \cite{nep_cpu}, e.g., via the interface provided by the \textsc{calorine} package \cite{calorine}, or on \glspl{gpu} by using the \textsc{gpumd} package.   
The \gls{sagpr} and \gls{tnep} (\gls{cpu}) models were tested on a server containing two Intel XEON Platinum 8275CL processors with a system memory of \qty{256}{\giga\byte}, while the \gls{tnep} (\gls{gpu}) models were tested on a heterogeneous server containing two Intel XEON Gold 6148 processors and an Nvidia GeForce RTX 4090 card with a graphics memory of \qty{24}{\giga\byte}.

\begin{figure}
\centering
\includegraphics{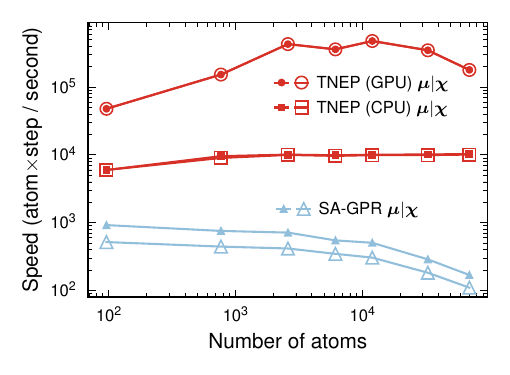}
\caption{
    Comparison of computational speed of \gls{sagpr} and \gls{tnep} models for dipole ($\vec{\mu}$) and susceptibility ($\tensor{\chi}$) of liquid water.
    Here, the \gls{sagpr} results were obtained using the \texttt{TENSOAP-FAST} implementation \cite{tensoap_fast}.
}
\label{fig:speed}
\end{figure}

The comparisons show that for system sizes $\gtrsim\,1000$ atoms the \gls{tnep} \gls{cpu} models are at least one order of magnitude faster than the \gls{sagpr} models on \glspl{cpu} for both dipole and polarizability (\autoref{fig:speed}).
On \glspl{cpu} the \gls{tnep} models exhibit nearly perfect weak scaling over the system sizes considered here.
In contrast, the \gls{sagpr} models show a notable decrease in speed as the system size increases.
Running the \gls{tnep} models on \glspl{gpu} enables an additional speed up by an order of magnitude or more.
For very small systems the \gls{gpu} implementation is limited by IO.
In addition we note that \textsc{gpumd} allows one to evaluate \gls{tnep} models on-the-fly during \gls{md} simulations for prediction of tensorial properties with a small impact on simulation speed (\autoref{sect:on-the-fly}).

\section{Applications}

Having established the accuracy and computational performance of the \gls{tnep} approach by comparison with reference data sets, we now demonstrate the application of \gls{nep} and \gls{tnep} models in combination for predicting \gls{ir} and Raman spectra of molecules, liquids, and solids.
To this end, we employ the correlation function approach outlined above (\autoref{sect:workflow} and \autoref{fig:workflow}).

\subsection{IR spectrum of water}

Firstly, we developed a \gls{nep} \gls{pes} model for liquid water using energy, atomic forces, and virial data from \gls{dft} calculations (\autoref{sect:pes-model-for-water}).

Next, a system of \num{216} water molecules was equilibrated in the NPT ensemble for \qty{100}{\pico\second} using the trained \gls{pes} model at \qty{298}{\kelvin} and \qty{1}{\bar}, followed by a further equilibration run in the NVT ensemble for another \qty{100}{\pico\second}.
Three production runs were carried out in the NVE ensemble for a duration of \qty{200}{\pico\second}.
A time step of \qty{0.5}{\femto\second} was used throughout.
We note that quantum effects can be actually rather pronounced in water as has been shown by path integral \gls{md} simulations in, e.g., Refs.~\citenum{KapWilLan20, SheLanWil21, KapKovCsa23}.
Here, we, however, decided to carry out classical \gls{md} simulations in order to enable a one-to-one comparison with the results of earlier studies.

The time dependence of the dipole ($\vec{\mu}(t)$) was computed for the production trajectories with a spacing of \qty{1}{\femto\second} using the \gls{tnep} dipole model for liquid water described above (\autoref{sect:dipole-model}).
The \gls{ir} spectrum was then obtained by Fourier transforming the dipole moment \gls{acf} via Eq.~\eqref{eq:ir-intensity}.
The final \gls{ir} spectrum was obtained by averaging the \gls{ir} spectra from the production runs.

For comparison, we also ran a \qty{200}{\pico\second} \gls{md} simulation with the TIP3P force field \cite{Praprotnik_water_ir_2005} via the CP2K software package \cite{Kühne_cp2k_2020}, where the TIP3P force field uses charges of \qty{-0.834}{\electron} and \qty{+0.417}{\electron} for oxygen and hydrogen, respectively.

\begin{figure}[b!]
  \centering
  \includegraphics{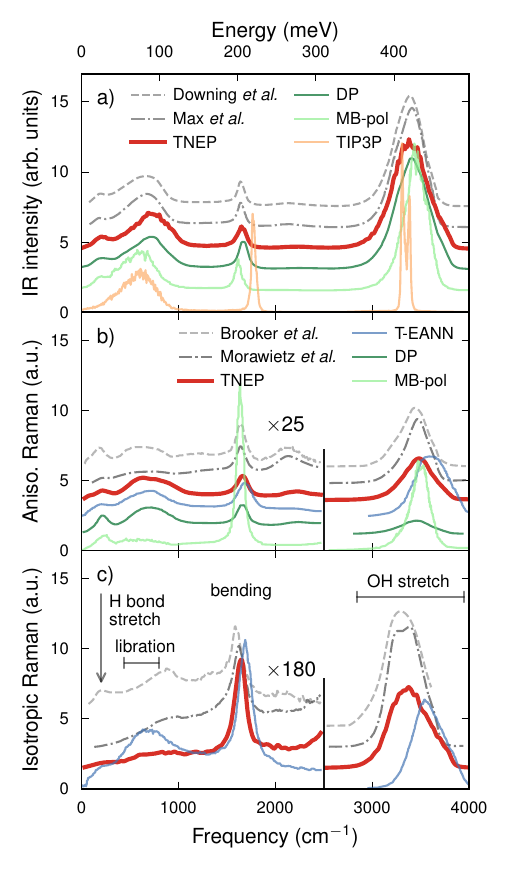}
  \caption{
    Comparison of (a) infrared as well as (b) anisotropic (depolarized) and (c) isotropic (polarized) Raman spectra of water at ambient conditions from simulations and experiment.
    Experimental data from Downing \textit{et al.} \cite{Downing1975}, Max \textit{et al.} \cite{Max2009_ir}, Brooker \textit{et al.} \cite{Brooker_exp_raman_1989}, and Morawietz \textit{et al.} \cite{morawietz_interplay_2018}. 
    Simulated spectra from \gls{teann} \cite{zhang_efficient_2020}, MB-pol \cite{Medders2015}, and \gls{dp} \cite{sommers_raman_2020, zhang_modeling_2021} models were adapted from the literature.
    In (a) and (b) the spectra were normalized by the integral between \num{80} and \qty{2500}{\per\centi\meter}, while in (c) they were normalized by the integral between \num{1000} and \qty{2500}{\per\centi\meter}.
  }
  \label{fig:water-spectra}
\end{figure}

The \gls{nep}-\gls{tnep} method yields an \gls{ir} spectrum that is in very good agreement with experimental data \cite{Downing1975, Max2009_ir} over the entire frequency range from \num{0} to \qty{4000}{\per\centi\meter} (\autoref{fig:water-spectra}a).
This includes the hydrogen-bond stretching band \cite{sommers_raman_2020} between \num{160} and \qty{250}{\per\centi\meter}, the libration band \cite{sommers_raman_2020} from \num{400} to \qty{800}{\per\centi\meter} associated with hindered molecule rotations \cite{Medders2015}, the bending modes \cite{Ceotto2021_solvate, Medders2015} at about \qty{1650}{\per\centi\meter} as well as the OH stretching band \cite{Ceotto2021_solvate, Medders2015} from \num{2800} to \qty{4000}{\per\centi\meter}.
The \gls{nep} and \gls{tnep} models for \gls{pes} and $\vec{\mu}$ in conjunction with the underlying exchange-correlation functional thus succeed in capturing the entire range stretching from the soft intermolecular to the stiff intramolecular modes.
This performance is also observed for the \gls{dp} model (\autoref{fig:water-spectra}a).

By comparison classical models produce rather large errors for the location of several features in the \gls{ir} spectrum of water.
\Gls{md} simulations with classical force fields \cite{praprotnik_temperature_2004, Praprotnik_water_ir_2005} such as TIP3P (\autoref{fig:water-spectra}a) and SPC/E tend to predict a blue-shifting of the bending modes by roughly \num{100} to \qty{200}{\per\centi\meter}.
A similar tendency was also observed for the POLI2VS model \cite{hasegawa_polarizable_2011}.
The results from the MB-pol model on the other hand exhibit a blue-shift of the OH stretching band by about \qty{50}{\per\centi\meter} \cite{Medders2015}.

The width of the OH stretching band has been proven to be quite difficult to predict due to the anharmonicity of the OH stretch mode \cite{Medders2015}.
The \gls{nep}-\gls{tnep} approach yields a value of \qty{380}{\per\centi\meter} for the full width at half maximum of this band, which is in good agreement with experimental estimates of about \qty{350}{\per\centi\meter} from Downing’s experiment \cite{Downing1975}.
Both \gls{nep}-\gls{tnep} and \gls{dp} predictions exhibit a slight high-frequency tail for this band, which is not visible in the experimental spectra.
This small difference could originate from the \gls{scan} functional \cite{Sun_scan_2015} that was used for generating the \gls{pes} training data \cite{sommers_raman_2020, Xu_dft_ir_2019} and/or the absence of  quantum effects in the (classical) \gls{md} simulations \cite{sommers_raman_2020, Medders2015}.  

\subsection{Raman spectra of water}

To obtain the Raman spectra of liquid water we sampled the time dependence of $\tensor{\chi}(t)$ using the \gls{tnep} susceptibility model and subsequently computed the \glspl{acf} for the same trajectories used for the prediction of the \gls{ir} spectra.
The full spectrum given by Eq.~\eqref{eq:raman-cross-section} and averaged over the available trajectories was then split into isotropic (polarized) and anisotropic (depolarized) contributions via Eq.~\eqref{eq:raman-lineshape-isotropic-anisotropic}.

The \textbf{anisotropic spectrum} predicted by the \gls{nep}-\gls{tnep} approach is overall in very good agreement with experimental data (\autoref{fig:water-spectra}b)  \cite{Brooker_exp_raman_1989, morawietz_interplay_2018}.
The locations of peaks and relative intensities of the stretching, bending, and librational modes in the simulated anisotropic Raman spectra are all well produced.
It is noteworthy that in the low frequency region below approximately \qty{1000}{\per\centi\meter}, the variation between the experimental spectra is larger than the variation between the \gls{ml} models and the experimental data.
This could be related to difficulties associated with processing the experimental raw data in this frequency region.

The \gls{teann} and \gls{dp} models yield similar results as the \gls{nep}-\gls{tnep} approach in the region up to about \qty{1900}{\per\centi\meter}.
On the other hand, all \gls{ml} models underestimate the intensity of the association band between \num{1900} and \qty{2500}{\per\centi\meter}, which is arising from the combination of librational and bending modes \cite{Medders2015, morawietz_interplay_2018}.
Here, the \gls{nep}-\gls{tnep} prediction is actually still the one that comes closest to the experimental spectra.

The broad high-frequency peak above \qty{3000}{\per\centi\meter}, which is associated with the OH stretch mode, is notably blue-shifted and broadened for the \gls{teann} model, while the \gls{dp} model strongly underestimates the intensity of this peak.
In contrast, the \gls{nep}-\gls{tnep} combination predicts this feature in good agreement with the experimental data.

Finally, the parametric MB-pol model yields the worst agreement with experiment, for example, strongly overestimating the intensity of the bending band while underestimating the libration band.

With regard to the \textbf{isotropic Raman spectrum} (\autoref{fig:water-spectra}c), one should first note the variation among the experimental data.
In particular in the region below \qty{1000}{\per\centi\meter}, the resulting uncertainty is comparable or even larger than the deviation between the \gls{nep}-\gls{tnep} prediction and the experimental data, while the position of the libration band predicted by \gls{teann} appears red-shifted.
With regard to the higher frequency region both \gls{nep}-\gls{tnep} and \gls{teann} reproduce the bending band well.
In the case of \gls{nep}-\gls{tnep} this also applies for the OH stretch band, whereas in the case of \gls{teann} a blue-shift can be observed similar to the anisotropic spectrum (\autoref{fig:water-spectra}b).

\subsection{IR spectrum of \texorpdfstring{\ce{PTAF^-}}{PTAF-}}
\label{sect:ptaf}

\begin{figure}[b!]
\centering
\includegraphics{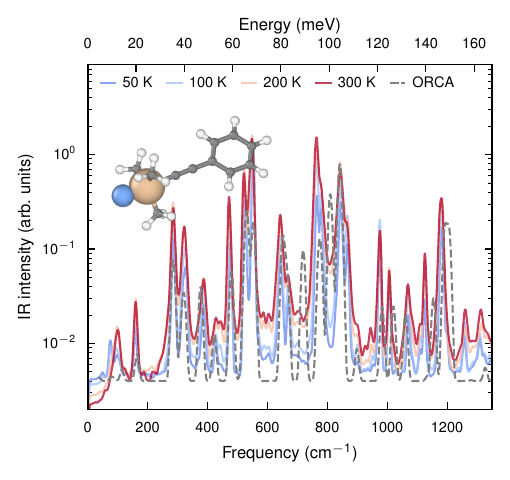}
\caption{
    \Gls{ir} spectra for the metastable \ce{PTAF^-} complex (see inset) at various temperatures.
    The gray dashed line represents the broadened integrated absorption coefficients of the harmonic spectrum obtained directly from \gls{dft} calculations.
    The overall agreement is good considering the lack of anharmonic corrections (intensity and vibrational frequencies) and temperature sensitivity of the spectrum obtained from \gls{dft} calculations.
}
\label{fig:ptafm}
\end{figure}

The \gls{nep}-\gls{tnep} method for predicting \gls{ir} spectra can be easily adopted for other molecular systems as long as the underlying observables to be learned are available.
Naturally, this includes the molecular configurations along a chemical reaction, such that experimentally observable spectral changes can be connected to metastable complexes.
One such complex is \ce{PTAF^-} (see inset in \autoref{fig:ptafm}), the intermediate reaction minimum in the  deprotection reaction 1-phenyl-2-trimethylsilylacetylene (PTA) with tetra-n-butylammonium ﬂuoride \cite{chintareddy2011tetrabutylammonium, thomas2016ground, schafer2022shining, Schafer_Fojt_Lindgren_Erhart_2024}.

To train \gls{nep} and \gls{tnep} models, we obtained \gls{pes} and ${\vec{\mu}}$ data for a set of \num{20170} structures via \gls{dft} calculations using the ORCA code \cite{neese2020orca}, the PBE functional \cite{Perdew_PBE_1996}, and a def2-TZVP basis set \cite{weigend2010segmented} while enforcing tight convergence of the self-consistent field cycles.
Subsequently, \gls{md} simulations at various temperatures were performed in the NVE ensemble using a timestep of \qty{0.1}{\femto\second} for \qty{1}{\nano\second}, during which 
$\vec{\mu}(t)$ was recorded with a time resolution of \qty{0.5}{\femto\second}.

The \gls{ir} spectra obtained via the analysis of the \gls{acf} of $\vec{\mu}$ show a pronounced temperature dependence in particular of the linewidths (\autoref{fig:ptafm}).
The molecule supports several soft modes with frequencies in the region below \qty{250}{\per\centi\meter}, which are associated with bending of and rotation about the ethynyl linker.
These modes in particular lead to strong mode coupling (i.e., anharmonicity), which underlies the changes in linewidth and the redistribution of the dipole strength across the spectrum.
Here, the computational efficiency of the \gls{nep}-\gls{tnep} implementation in \textsc{gpumd} was crucial to resolve these features, as it enabled sampling on the nanosecond time scale, which would be prohibitive for a \gls{dft}-\gls{md} simulations and computationally very expensive for a \gls{cpu} implementation.

\subsection{Raman spectra of \texorpdfstring{\ce{BaZrO3}}{BaZrO3}}

\begin{figure}[b!]
\centering
\includegraphics{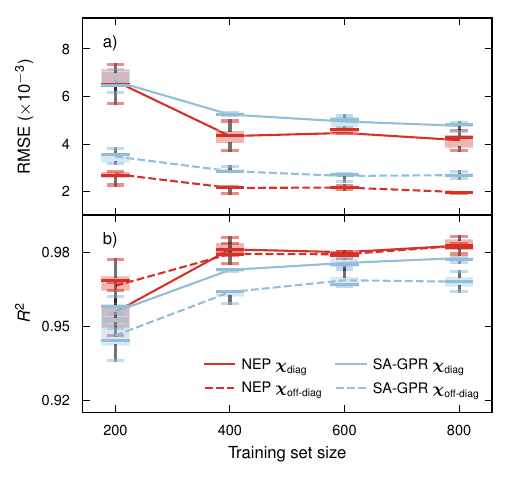}
\caption{
    Variation of (a) \glspl{rmse} and (b) $R^2$ scores with training set size for \gls{tnep} and \gls{sagpr} models based on five training sets per size generated by shuffle-split.
    In the case of \gls{tnep}, we used $N_\text{neu}=20$, $n_\text{max}^\text{R}=n_\text{max}^\text{A}=4$, and $\lambda_1=\lambda_2=\num{2e-3}$ (compare Figs. \ref{fig:BZO-chi-architecture} and \ref{fig:BZO-chi-lambda-convergence}).
}
\label{fig:BZO-chi-size-convergence}
\end{figure}

\begin{figure}
\centering
\includegraphics{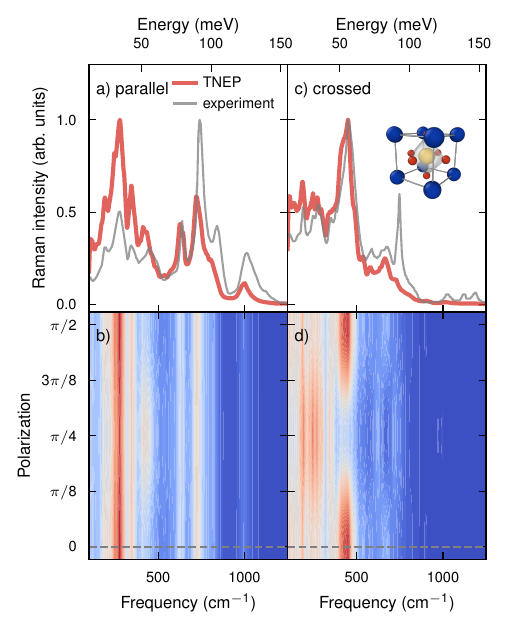}
\caption{
    Raman spectra of \ce{BaZrO3} for (a,b) parallel and (c,d) crossed polarization from simulations using a combination of \gls{nep} and \gls{tnep} models (red lines) as well as experiment (gray lines) \cite{TouAmo2019}.
    The spectra shown in (a,c) have been predicted for the nominal alignments used in the experimental measurements.
    The corresponding polarizations are indicated by the dashed horizontal lines in (b,d).
}
\label{fig:BZO-raman}
\end{figure}

\ce{BaZrO3} is a perovskite that is being investigated, e.g., as a proton conductor for applications in fuel cells.
It has also been the subject of various fundamental studies, as it is a prototypical antiferroelectric perovskite \cite{Akbarzadeh2005, Knight2020, PerJedRomPioLinHylWah20}.
It features soft and strongly temperature-dependent phonon modes \cite{RosFraBrauTouBouAndBosMaeWah23, FraRosErh23}, which have been carefully analyzed with Raman spectroscopy \cite{TouAmo2019}, rendering \ce{BaZrO3} an ideal application for the present approach.

For benchmarking, we constructed models for $\tensor{\chi}$ using both the \gls{tnep} and \gls{sagpr} approaches.
The reference data set comprised cubic and tetragonal supercells with up to \num{40} atoms.
The training structures were taken from \gls{md} simulations at different temperatures and pressures, generated using a \gls{nep} \gls{pes} model constructed in an earlier study \cite{FraRosErh23}.
In total the reference data set contained \num{940} structures.
\num{140} structures were randomly placed in a hold-out set for validation, while training sets were compiled by the shuffle-split method (random selection with replacement) with \num{200} to \num{800} structures and five data sets per training set size.

A comparison of models generated using different choices for the size of the neural network as well as the descriptor demonstrates that viable models can be obtained for a wide range of parameters, and that even small models with as few as \num{1500} or so parameters can yield very good results (\autoref{fig:BZO-chi-architecture}).
Yet fine-tuning of these parameters as well as the regularization parameters (\autoref{fig:BZO-chi-lambda-convergence}) allows one to maximize model performance.

The convergence of \glspl{rmse} and $R^2$ scores with training set size is similar for \gls{tnep} and \gls{sagpr} with a slightly better performance for \gls{tnep} (\autoref{fig:BZO-chi-size-convergence}).
In both cases, training sets of about \num{400} structures already yield very good models, demonstrating the data efficiency of these approaches.
This behavior has also been observed in the construction of models for amino acids \cite{BerNieLam24}.

Next \gls{md} simulations were carried out using \numproduct{12x12x12} supercells (\num{8640} atoms) and a timestep of \qty{1}{\femto\second} using the \gls{nep} model for the \gls{pes}.
Following equilibration at \qty{300}{\kelvin} and \qty{0}{\giga\pascal} in the NPT ensemble, the time-dependent susceptibility $\tensor{\chi}(t)$ was recorded for \qty{500}{\pico\second} using a time resolution of \qty{5}{\femto\second}.
For production, we used a \gls{tnep} model for $\tensor{\chi}$ trained against the full data set but we found that models based on at least approximately \num{400} structures to yield results that are practically indistinguishable within the statistical uncertainty.
The Raman lineshape was subsequently obtained via the \gls{acf} of $\tensor{\chi}$ according to Eq.~\eqref{eq:raman-lineshape}.
We then computed the Raman spectra for parallel (\autoref{fig:BZO-raman}a,b) and crossed polarization (\autoref{fig:BZO-raman}c,d), which in Porto notation correspond to $Z(XX)\bar{Z}$ and $Z(XY)\bar{Z}$, respectively, where $X$ and $Y$ are arbitrary crystal axes.
The final spectra were obtained by averaging over \num{20} independent \gls{md} trajectories.

The results are overall in very good agreement with experiment, especially considering the very strong anharmonicity of this material and the strong temperature dependence of the vibrational spectrum \cite{RosFraBrauTouBouAndBosMaeWah23}.
The main difference with respect to the position of the peaks is a slight red-shift in the predicted spectra in the region above \qty{600}{\per\centi\meter}.
This overly soft response can be attributed to the underlying exchange-correlation functional (vdW-DF-cx, Refs.~\citenum{DioRydSch04, BerHyl2014}), which the \gls{nep} model truthfully reproduces.
One can also observe an inversion in the intensity of the low and high energy features.
This effect is almost certainly due to the classical sampling used here.
It is rather common to correct for quantum effects in \gls{ir} and \emph{first order} Raman spectra by including a factor similar to the prefactor in Eq.~\eqref{eq:ir-intensity}.
In the case of \ce{BaZrO3} the room-temperature Raman spectrum arises, however, due to \emph{second-order} scattering, i.e., due to combinations of modes.
In that case, the application of the commonly used correction factor is no longer valid.
Here, we therefore omit such corrections entirely.

The Raman spectra depend on the crystal orientation with respect to the excitation laser.
The present approach allows one to readily map out this dependence via Eqs.~\eqref{eq:raman-cross-section} and \eqref{eq:raman-lineshape} (\autoref{fig:BZO-raman}b,d).
While we are unaware of experimental measurements of the polarization dependence for \ce{BaZrO3}, we note that such experiments have been carried out for, e.g., \ce{NaCl} \cite{Benshalom_Reuveni_Korobko_Yaffe_Hellman_2022}.
As demonstrated in the former study, such measurements can provide valuable additional information.

\section{Conclusions}

In this contribution, we have introduced an extension of the \gls{nep} approach to tensors, resulting in the \gls{tnep} scheme.
This was achieved by constructing expressions for rank-1 and rank-2 tensors based on the expression for the virial, which is a rank-2 tensor that arises naturally from derivatives of the energy (a rank-0 tensor) with respect to the atomic distances.
This approach, which can be extended to tensors of higher rank, thus allows one to easily construct models that are equivariant.

We demonstrated the accuracy of this approach and its computational efficiency by constructing models for the dipole moment $\vec{\mu}$, the molecular polarizability $\tensor{\alpha}$, and the electric susceptibility $\tensor{\chi}$ for several molecules, a liquid as well as two crystalline materials.
In particular, the computational speed of the current method and its implementation in the \textsc{gpumd} package provide a significant advantage both in terms of the time scales and system sizes that can be sampled.

Finally, we applied the approach to predict \gls{ir} and Raman spectra of liquid water, the molecule \ce{PTAF^-}, and the perovskite \ce{BaZrO3} in very good agreement with available experimental data, illustrating the range of systems that can be readily addressed using the \gls{tnep} methodology introduced here.

\section*{Acknowledgments}

N.~X. is grateful for the financial support provided by the Startup Funds of the Institute of Zhejiang University-Quzhou.
P.~R., C.~S., E.~L., N.~Ö., and P.~E. acknowledge funding from the Swedish Research Council (Nos.~2020-04935 and 2021-05072), the European Union under the Marie Sk{\l}odowska-Curie program (No.~101065117), and the Swedish Foundation for Strategic Research via the SwedNESS graduate school (GSn15-0008) as well as computational resources provided by the National Academic Infrastructure for Supercomputing in Sweden at NSC, PDC, and C3SE partially funded by the Swedish Research Council through grant agreement No.~2022-06725.
Y.~H. acknowledges funding from the National Key Research and Development Program of China (Nos. 2022YFE0106100), and the National Natural Science Foundation of China (Nos. 22178299, 51933009).
W.~C. acknowledges funding from the National Natural Science Foundation of China (Nos. 22373104).

\section*{Supporting information}

\textbf{Supplemental material.}
Training process of \gls{sagpr} models for water systems; calculations of dipole moment for liquid water and $\alpha$-\ce{Fe2O3}; notes on the Clausius-Mossotti relation and the units of polarizability and electric susceptibility; parity plots of the \gls{tnep} predicted dipole moment, diagonal and off-diagonal elements of polarizability versus the \textit{ab initio} references for water systems; demonstration of rotational invariance; parameters used in the training of \gls{nep} and \gls{tnep} models; dependence of \gls{tnep} model performance for \ce{BaZrO3} on model and training hyperparameters; timing of on-the-fly model evaluation during \gls{md} simulations.

\textbf{Data availability.}
The source code and documentation for \textsc{gpumd} are available at \url{https://github.com/brucefan1983/GPUMD} and \url{https://gpumd.org}, respectively.
The source code and documentation for \textsc{calorine} are available at \url{https://gitlab.com/materials-modeling/calorine} and \url{https://calorine.materialsmodeling.org}, respectively.
\Gls{nep} models and data are available via Zenodo via
\url{https://doi.org/10.5281/zenodo.10257363} (water and \ce{BaZrO3}),
\url{https://doi.org/10.5281/zenodo.10255268} (\ce{PTAF^-}), and
\url{https://doi.org/10.5281/zenodo.8337182} (\ce{BaZrO3}).

\end{document}